# An Indoor Crowd Movement Trajectory Benchmark Dataset

Ying Zhao, Xin Zhao, Siming Chen, Zhuo Zhang, and Xin Huang

*Abstract*—In recent years, technologies of indoor crowd positioning and movement data analysis have received widespread attention in the fields of reliability management, indoor navigation, and crowd behavior monitoring. However, only a few indoor crowd movement trajectory datasets are available to the public, thus restricting the development of related research and application. This paper contributes a new benchmark dataset of indoor crowd movement trajectories. This dataset records the movements of over 5000 participants at a three-day large academic conference in a two-story indoor venue. The conference comprises varied activities, such as academic seminars, business exhibitions, a hacking contest, interviews, tea breaks, and a banquet. The participants are divided into seven types according to participation permission to the activities. Some of them are involved in anomalous events, such as loss of items, unauthorized accesses, and equipment failures, forming a variety of spatial–temporal movement patterns. In this paper, we first introduce the scenario design, entity and behavior modeling, and data generator of the dataset. Then, a detailed ground truth of the dataset is presented. Finally, we describe the process and experience of applying the dataset to the contest of ChinaVis Data Challenge 2019. Evaluation results of the 75 contest entries and the feedback from 359 contestants demonstrate that the dataset has satisfactory completeness, and usability, and can effectively identify the performance of methods, technologies, and systems for indoor trajectory analysis.

## I. Introduction

HUMAN daily activities generate massive amounts of crowd trajectory data in either indoor or outdoor environments [1]–[18]. GPS (global position system) technologies based on global positioning satellites support the collection of outdoor trajectory data. Short-distance positioning technologies, such as RFID, Bluetooth, UWB, and WLAN [19]–[25], are the main approaches to obtain indoor trajectory data because GPS signals will be interfered with or blocked by buildings. The increase in crowd trajectory data availability offers new opportunities of analysis and assessment for reliability engineering [26], [27]. Analysts can discover human movement patterns and behavior characteristics in indoor scenes, such as offices, subways, hospitals, shopping malls, and meeting halls, by using indoor trajectory data, which enables reliable and humanized services (e.g., anomaly detection, indoor navigation, and hotspot recommendation [28]–[34]) in building management. In particular, amid the recent COVID-19 pandemic, indoor trajectory data have shown their considerable research value for epidemic prevention and control [35]–[38].

High-quality data are vital to boost the development of data science. In recent years, numerous open-source datasets of outdoor crowd trajectories, such as transportation management, population mobility management, and safety monitoring of major incidents, have empowered the rapid development of related fields [39]–[52]. However, open-source indoor crowd trajectory datasets are currently rare mainly due to the following three reasons. (1) Short-range positioning technologies are not as mature and popular as GPS. (2) Indoor scenes are more private and sensitive and easily trigger privacy problems compared with outdoor scenes. (3) Providing complete, customized, and vivid stories in real-world indoor crowd trajectory data is difficult.

Therefore, creating and using simulation data is a feasible solution. Simulation data are not restricted to hardware devices of indoor trajectory data collection. These data do not trigger privacy problems and can be embedded in dramatic stories, thereby fulfilling most research requirements. However, creating high-quality indoor crowd trajectory data is difficult. The challenges mainly come from three aspects. (1) The movements of each individual in the real world are affected by various objective and subjective factors, such as physical strength, current locality, crowd density, and subjective interest. (2) Indoor environments generally have compact structures, diverse compositions, and complex functionalities. (3) The crowd distribution in an indoor space is constantly changing. Thus, such changes must conform to the structures and functionalities of the physical indoor space.

This paper introduces an indoor crowd trajectory benchmark dataset called Indoor Crowd Movement Trajectory Dataset 2019 (ICMTD-2019) [53]. The scene of the dataset was set at a fictitious international cyber security academic conference. The conference, which lasted for three days with more than 5000 people comprised three major parts, namely academic seminars, business exhibitions, and hacking contests. Moreover, a series of social activities, such as tea breaks, interviews, and a banquet, were conducted. Each participant or organizer wore a smart badge to capture his/her movements in the public areas of the venue during the conference. Consequently, the ICMTD-2019 dataset included the following: (1) trajectories of 5256 people in the venue during the three days, the conference schedule, and venue map (refer to the supplementary materials of the paper); (2) varied movement patterns of various types of participants as the conference schedule proceeds as planned; (3) diverse abnormal events, such as equipment failure, loss of items, personnel ultra vires, and congestion in the venue, which often lead to troubles to participants and organizers. All these movement patterns and abnormal events were comprehensively presented in the ground truth (refer to the supplementary materials of the paper).

This paper presents the comprehensive process of modeling and generation of the dataset. First, the scene is determined through on-site investigations on a real-world large convention center and in-depth interviews with experienced academic conference organizers. Then, models are established to describe three main types of entities in the scene, namely characters, activities, and locations. Next, methods to drive, constrain, and control the behaviors of each character are proposed. Finally, a program-driven data generation tool is designed to generate the ICMTD-2019 dataset.

The dataset had been evaluated by an established data challenge named the 2019 ChinaVis Data Challenge [54], [55]. The challenge attracted 359 contestants and received 75 submissions. The analysis difficulties of movement patterns and abnormal events contained in the dataset were examined on the basis of these submissions. The feedbacks were collected from the returned questionnaires from the contestants to evaluate the quality, usability, and rationality of the dataset. The results showed that the ICMTD-2019 dataset had a complete scene design and a reasonable and realistic modeling of characters and activities in the scene. The various spatial–temporal movement patterns and vivid storylines of the dataset stimulated the enthusiasm of the contestants for data analysis and helped identify the effectiveness of data analysis methods, technologies, and systems employed by the contestants. Analyzing user behaviors in the academic conference as a complex system is vital for reliability and safety management. Especially, with the proposed open-sourced data, more algorithms can be tested and evaluated for pattern seeking and anomaly detection [56]–[58]. Researchers and practitioners in reliability management can be benefit from the situation awareness of patterns and anomalies [59], [60].

Overall, this paper provides the following contributions: (1) an indoor crowd movement trajectory benchmark dataset, (2) a set of indoor crowd movement behavior modeling and data generation methods, and (3) a series of experiences in using the dataset to evaluate approaches on indoor crowd movement trajectory analysis.

## II. Related Work

### A. Previous trajectory benchmark data

Numerous outdoor trajectory benchmark datasets are available due to their mature outdoor positioning technologies and developed relevant research fields. For example, vessel trajectories implied in Automatic Identification Systems (AIS) data are crucial to obtain a good understanding of the maritime traffic situation and reduce the hazards of marine navigation [1]. The EPFL Mobility [2], [3], and T-Drive taxi trajectories datasets [4], [5], are high-precision GPS data of taxis that are widely used in traffic flow analysis and traffic network optimization. The basketball shot image dataset [6] is used to analyze the correctness of basketball shot trajectory, which has good significance in guiding basketball training. The USCD [7] and WorldExpo'10 datasets [8], which provide trajectory data of pedestrians captured by surveillance cameras, have been extensively used for outdoor crowd behavior analysis.

At present, few indoor trajectory benchmark datasets are available due to the immaturity of indoor positioning technologies, complex and compact structures of indoor spaces, and susceptibility to privacy issues. The Mall dataset [10]–[13], a real indoor trajectory dataset, provides the movement information of 60,000 people captured by surveillance cameras in a shopping mall. However, it is hard for real-world data to include complete storylines with diverse events.

IEEE VAST Challenge 2016 dataset is a synthetic indoor trajectory dataset highly related to our dataset [14], [15]. It offers employee trajectories and environment sensing data (e.g., concentration levels of carbon dioxide and hazium) in an office building. It can be used to investigate indoor crowd behaviors and their correlations with indoor environment shifts. Compared to IEEE VAST Challenge 2016 dataset, the first strength of our dataset is that our venue map is fine-grained with $8*8m^2$ grids whereas the building map of IEEE VAST 2016 dataset is divided by coarse-grained functional areas. Second, our dataset involves more than 5,000 people but IEEE VAST Challenge 2016 dataset only has 125 employees. Third, our scenario includes a vivid storyline with 34 parallel conference activities and 12 anomalous events.

### B. Trajectory data generation

The generation methods of simulated trajectory data can be roughly divided into data- and program-driven methods. Data-driven methods generate simulated trajectories in scenes similar to those of existing trajectory data. For example, Lerner *et al*. [61] extracted the trajectories of moving objects from videos and then generated new trajectories based on the extracted ones. Lee *et al*. [62] built an agent model based on the crowd trajectories obtained from videos and then generated trajectories of a virtual crowd. Pappalardo *et al*. [63] constructed a trajectory simulator to learn patterns hidden in various movement logs to generate simulated trajectories. Implementing data-driven methods is easy, and generated trajectories contain the movement patterns of referenced real moving objects. However, the diversity of patterns and scenes is restricted by references, and the expandability of the data patterns is limited.

Program-driven methods generate trajectories of moving objects by designing models, algorithms, or tools. Compared with data-driven methods, program-driven approaches can independently construct scenes and freely control data complexity. However, the difficulty of program-driven methods mainly lies in ensuring the closeness of the simulated scenes and movement patterns to the real world. Jensen *et al*. [64], [65] used a random algorithm to identify the next position of a moving object and then prompted the object to move to a random target position through the Dijkstra algorithm.

However, the random algorithm does not consider the moving characteristics of the object. Huang et al. [66] materialized a trajectory data generation tool, namely IndoorSTG, which defines the degree of interest of a moving object in each spatial position to determine its next position. However, this tool is limited to produce diverse movement patterns. Li et al. [67] built a tool named Vita to generate multiple types of indoor trajectories in a certain building by using distribution control models of movements. Users can customize the number of moving objects, their maximum speed, and life cycles by adjusting parameters. However, the tool cannot guarantee the structures and functionalities of complex indoor spaces, and the generated data are not embedded into vivid story plots.

The ICMTD-2019 dataset introduced in this paper is generated by a program-driven method. Compared with the above-mentioned indoor crowd trajectory generation methods, the proposed method considers various objective and subjective factors that drive movements of people, builds a set of entity-representation and behavior-constraint models to ensure the diversity and reality of movement patterns, and embeds a variety of scheduled social activities and unexpected events into the data, to the best of our knowledge, which have not been previously conducted.

## III. Scenario Design

Four basic principles are identified in this paper to guide the scene definition. (P1) Realistic: the scene must be close to the real world although it is fictional; (P2) Universality: the scene must be universal and easy to understand; (P3) Diversity: the scene must contain a wealth of entity objects, events, and storylines; (P4) Challenging: the scene must be relatively difficult to stimulate the enthusiasm for analysis and promote the research of relevant topics.

The scene of the ICMTD-2019 dataset is set in a large international cybersecurity academic conference. Multiple real-world cyber security academic conferences are used as a reference to ensure realism and universality (P1 and P2), and coordination is made with one venue manager and two academic conference organizers during the entire scene design process. Seven types of personnel, including VIP guests, ordinary guests, visitors, media reporters, hacking contest participants, staff, and exhibitors, are designed with specific participation permissions and different movement patterns to ensure diversity (P3). A two-story indoor venue space, which can accommodate more than 5000 people, is also designed on the basis of the actual venue of a real-world large international conference center. Space contains the main venue, a sub-venues area, an exhibition area, a contest area, and other functional areas, including service desks, canteens, coffee break areas, leisure areas, and restrooms, as well as other rooms for related types of personnel, such as VIP lounges, media rooms, and work rooms, to rest and work. Two real-world international cyber security academic conferences are considered to design many parallel conference activities, such as keynotes, academic seminars, business exhibitions, a hacking contest, and symposiums. Twelve abnormal events are designed to increase difficulty (P4). Some of these events involve fine-grained spatial–temporal patterns and complex correlations, which are difficult to detect. Thus, the proposed dataset can be used to evaluate the capability and effectiveness of methods, technologies, and systems used by data analysts.

TABLE I
DESCRIPTION OF THE CONFERENCE PERSONNEL MODEL

| Type | Attributes | Description | Range |
|---|---|---|---|
| Basic Attributes | PID | Conference personnel number | - |
| | Gender | Gender | - |
| | Age | Age | - |
| | Job | Occupation | - |
| | Type | Personnel type | 0, 1, 2, … ,6 (Section VII) |
| Status Attributes | MS | Motion status | [Moving, Stay] |
| | BS | Behavior status | [Absent, Focused, Occupied] |

## IV. Entity Modeling

Three entity models are designed to describe the characters of people, conference activities, and venues (indoor spaces in the scene). These models are comprehensively explained below.

### A. Conference personnel model

The conference personnel model describes each individual in the scene with five basic attributes and two status attributes, as shown in Table I. The basic attributes, which have fixed values during the conference, specify the identity, background, and type of an individual. Seven personnel types are pre-defined, and each type has its participation permissions (Section VII). The status attributes depict the motion and behavior statuses of an individual at a time, updating once per second by default. Two motion statuses are available: moving and stay. Three behavior statuses are also available: not in the venue (Absent); in the venue and actively participating in a certain conference activity (Focused), such as attending an academic seminar out of personal interests; passively participating in a certain conference activity (Occupied), which generally occurs in conference personnel with specifically assigned tasks, including seminar chairs and keynote speakers.

### B. Conference activity model

A conference schedule, which has 12 academic activities in the main venue, 22 academic activities in the sub-venues, and a hacking contest, is designed with reference to two real-world cyber security academic conferences. The schedule also includes exhibitions, tea breaks, lunches, media interviews, and a banquet. The detailed schedule is presented in the supplementary materials of this paper. A conference activity model is designed to describe each activity from four aspects, namely basic, permission, priority, and status attributes, as shown in Table II. The basic attributes depict the name, start time, end time, and location of an activity, which are fixed during the conference. The permission attribute indicates the types of conference personnel that is allowed to participate in an activity.

For example, exhibitions allow all types of personnel to visit,

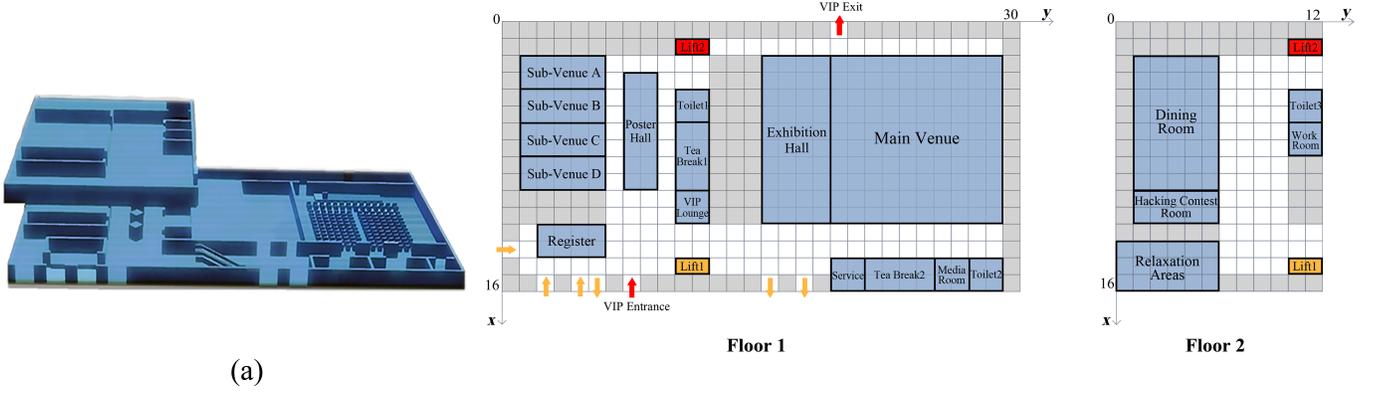

Fig. 1. Indoor structure of conference venue: (a) 3D model; (b) 2D top view and functional zones

TABLE II
DESCRIPTION OF THE CONFERENCE ACTIVITY MODEL

| Type | Attributes | Description | Range |
|---|---|---|---|
| Basic Attributes | AName | Name of conference activity | - |
| | $T_{start}$ | Start time | - |
| | $T_{end}$ | End time | - |
| | Location | Location | - |
| Permission Attribute | Permission | Personnel type allowed to attend | Conference personnel types |
| Priority Attribute | Pr | Priority | 0, 1, 2, … ,6 |
| Status Attribute | EP | Completion progress | [0,1] |

TABLE III
DESCRIPTION OF THE VENUE SPACE MODEL

| Type | Attributes | Description | Range |
|---|---|---|---|
| Basic Attributes | X | Abscissa | [0, 30] |
| | Y | Ordinate | [0, 16] |
| | Floor | Floor | [floor1, floor2] |
| Capacity Attributes | Speed | Max moving speed | 0–4 m/s |
| | Capacity | Max capacity | 0–400 |
| Status Attributes | Number | Current capacity | 0–400 |
| | StayList | Current list of personnel | - |

whereas the hacking contest only permits the attendance of contestants and staff. The priority attribute represents the priority of each conference activity. For example, academic activities are generally prioritized over exhibitions to attract conference personnel. The status attribute outlines the completion progress of an activity, which is determined by the current time, start time and end time of an activity.

*C. Conference venue model*

Fig. 1 shows the 3D model and the 2D top view of the indoor space of the conference venue. An area of 8 × 8 square meters is used as a space unit to divide the venue space considering the typical communication distance (1–15m) of UHF-RFID positioning devices [68]. Fig. 1(b) shows that the indoor space of the venue is equally divided into non-overlapping grids. Where the grids marked in red represent the special access for VIPs. Therefore, coordinates can be used to depict the spatial structure and functional zones of the venue, and also to facilitate the description of the trajectory.

A space model is built to describe the basic, capacity, and status attributes of a space unit, namely a grid, as shown in Table III. The basic attributes refer to the location information of a grid. The capacity attributes describe the maximum number of people that a grid can accommodate and the max speed that allows people to pass through. A grid is stipulated to accommodate 0 to 400 people with an allowable speed of 0 to 4 m/s after onsite investigations and measurements. The capacity values of grids in different functional zones are set differentially. For instance, the capacity is high for the corridor, but low for the VIP lounge. In addition, the diversity of capacity values leads grids in different functional areas to have different max moving speeds. The larger the capacity in a grid, the lower the max moving speed. Status attributes, including the real-time capacity and personnel list, depict the current accommodation status of a grid with an updating frequency per second.

## V. BEHAVIOR MODELING

People at the conference will participate in various activities in different areas of the building. Such participation behaviors are modeled in this section. Our model design draws on the experience of venue managers and conference organizers as well as the existing researches on crowd modeling, crowd simulation, and behavior modeling [70]–[74].

*A. Behavior-constraint model*

The behavior-constraint model describes four objective factors that can affect crowd movements in the venue, as shown in Table IV. Time constraints define the venue opening hours of three days. Conference personnel can only move to the venue during the opening hours. Permission constraints define the types of personnel that are allowed to enter a certain functional zone in the venue. For example, VIP lounge only allows VIP guests to enter.

Energy constraints limit how long people can continually stay in the venue. E represents the current energy value of an individual. The energy value of an individual is upon daily entrance to the venue, and the value gradually decreases as the time of stay increases. An energy constraint function is defined to simulate the change in E, as shown in Formula 1:

$$E_{i,t} = (1 - EDR_i)(E_{i,t-1} - E_{min_i}), \qquad (1)$$

where $E_{i,t}$ and $E_{i,t-1}$ represent the energy value of person $i$ at time $t$ and $t$-1, respectively, $EDR_i$ is the energy decay rate of person $i$, and $E_{min_i}$ is the minimum energy value of person $i$. This formula shows that $E_{i,t}$ is gradually close to the minimum energy value $E_{min_i}$ under the influence of the $EDR_i$. Moreover, an exit target value $ET$, which is usually slightly larger than $E_{min}$, is set for an individual. A lower $E_i$ than the $ET_i$ means that the energy of person $i$ is insufficient, prompting person $i$ to leave the venue. The initial $E_{min}$, $EDR$, and $ET$ of each conference personnel are randomly generated based on their age; for example, a young person generally has a slow energy decay rate and a small $E_{min}$.

TABLE IV
DESCRIPTION OF THE BEHAVIOR-CONSTRAINT MODEL

| Type | Attributes | Description | Range |
|---|---|---|---|
| Time Constraints | T1 | Opening hours on day1 | 7:30–18:00 |
| | T2 | Opening hours on day2 | 8:00–19:30 |
| | T3 | Opening hours on day3 | 7:30–13:00 |
| Permission Constraints | PAR | Personnel authority restriction | - |
| Energy Constraints | E | Energy value | [0, 1] |
| | $E_{min}$ | Min energy value | [0, 1] |
| | EDR | Energy decay rate | [0, 1] |
| | ET | Exit target value | [0, 1] |
| Capacity Constraints | SC | Saturation of capacity | [0, 1] |
| | SR | Speed ratio | [0, 1] |

TABLE V
DESCRIPTION OF THE BEHAVIOR-INTEREST MODEL

| Type | Attributes | Description | Range |
|---|---|---|---|
| Personal Interest Features | CIV | Current interest value | [0, 1] |
| | IDR | Interest decay ratio | [0, 1] |
| | IRR | Interest recovery ratio | [0, 1] |
| | $IV_{min}$ | Lowest interest | [0, 1] |
| | $IV_{max}$ | Highest interest | [0, 1] |
| Activity Attractiveness Features | CAV | Current attracting value | [0, 1] |
| | ADR | Attractiveness decay ratio | [0, 1] |
| | ADT | Attractiveness decay trend | [−1, 1] |

Capacity constraints control the number of people and their moving speeds in a space unit. The real-time crowd saturation $SC$ of a grid is the ratio of the current capacity to the maximum capacity. The entrance of additional people is prohibited when the saturation is equal to 1. The real-time speed ratio ($SR$) of a space unit is the ratio of the currently allowed maximum moving speed to the pre-defined maximum moving speed. The currently allowed maximum moving speed is 4 m/s. A speed constraint function is defined to control the change in $SR$, as shown in Formula 2:

$$SR_{x,y,z,t} = \cos\left[2\pi\, SC_{x,y,z,t} / (SC_{x,y,z,t}+1)^2\right], SC \in [0,1] \quad (2)$$

where $SR_{x,y,z,t}$ and $SC_{x,y,z,t}$ respectively represent the speed ratio and crowd saturation of a space unit located on coordinate $x, y$ of $z$-floor at time $t$. This formula shows that $SR$ gradually decreases with the increase in $SC$ in the monotonic interval of the cosine function. The product of $SR$ and $Speed$ is the current speed of a person.

*B. Behavior-interest model*

The behavior-interest model describes the subjective factors that drive crowd movements in the venue. The subjective factors in this scene fall into two main aspects, namely personal interest and activity attractiveness features, as shown in Table V. Personal interest features determine the real-time interest of an individual in a certain conference activity. These features comprise the following: *CIV*, which depicts the current interest value; interest decay ratio (*IDR*) and interest recovery ratio (*IRR*), which control the change in *CIV*; and $IV_{min}$ and $IV_{max}$, which define the value range of *CIV*. Activity attractiveness features determine the real-time attracting value of an activity to an individual. These features comprise *CAV*, which depicts the current attracting value, an *ADR* and *ADT*, which control the change in *CAV*. The *IDR*, *IRR*, $IV_{min}$, and $IV_{max}$ of each individual to each activity and the *ADR* and *ADT* of each activity to each individual are preset in accordance with the basic attributes of each individual and activity. That is, people in the venue have diverse preferences for conference activities.

The product of *CIV* and *CAV* is defined as the interest matching value (*IMV*) of an individual to an activity. That is, *CIV* and *CAV* jointly motivate an individual to decide activity switching. Two calculation methods are designed to simulate the real-time changes in *CIV* and *CAV*.

(1) *CIV* calculation

People generally tend to be curious of the unfinished conference activities that they have not yet participated in. Therefore, the *CIV*s of an individual to such activities are stipulated to increase gradually. Formula 3 provides the calculation method of such *CIV*.

$$CIV_{i,j,t} = (IRR_{i,j} - 1)(CIV_{i,j,t-1} - IV_{max_{i,j}}), \qquad (3)$$

where $CIV_{i,j,t}$ and $CIV_{i,j,t-1}$ respectively represent the *CIV*s of participant $i$ to to-be-attended activity $j$ at time $t$ and $t$-1, $IRR_{i,j}$ refers to the interest recovery ratio of participant $i$ to activity $j$, and $IV_{max_{i,j}}$ represents the highest interest of participant $i$ to activity $j$. For example, the interest of a person in the main venue to the exhibition would gradually increase in a non-linear manner before reaching the preset maximum.

Moreover, feelings of people to the activity that they are currently participating are categorized into two: gradual loss of interest or consistent high interest. Thus, the *CIV*s of an individual to the currently-participating activity may gradually decline, remain the same, or even rise. The attractiveness decay trend (*ADT*) is designed to simulate such phenomena. Formula 4 offers the calculation method of such *CIV*.

$$CIV_{i, current, t} = (IDR_{i, current} - 1)(1 - ADT_{i, current})$$

$$\times (CIV_{i, current, t-1} - IV_{min_{i, current}}), \qquad (4)$$

where $CIV_{i, current, t}$ and $CIV_{i, current, t-1}$ respectively represent the $CIV$s of participant $i$ to the currently-participating activity at time $t$ and $t$-1, $IDR_{i, current}$ is the interest decay ratio of participant $i$ to the currently-participating activity, $IV_{min_{i, current}}$ represents the lowest interest of participant $i$ on the activity, and $ADT_{i, current}$ refers to the attractiveness decay trend of participant $i$ to the activity. Notably, $ADT_{i, current}$ can be negative; that is, participant $i$ indicates enjoyment to the current activity and willingness to stay until the end.

(2) *CAV* calculation

The attractions of conference activities can be affected by their progresses. A conference activity at the beginning is generally more appealing to people who have not been involved than the activity near the end. Formula 5 offers the calculation method of such *CAV*.

$$CAV_{j, i, t} = CAV_{j, i, t-1}(1 - ADR_{j, i})[1 - \frac{(t - T_{start_j})}{(T_{end_j} - T_{start_j})}], \quad (5)$$

where $CAV_{j, i, t}$ and $CAV_{j, i, t-1}$ respectively represent the *CAV* of participant $i$ to activity $j$ at time $t$ and $t$-1, $(t - T_{start_j})/(T_{end_j} - T_{start_j})$ is the progress of activity $j$, and $ADR_{j, i}$ represents the attractiveness decay ratio of activity $j$ to participant $i$.

Moreover, an activity is growing less intriguing for people already present. Formula 6 offers the calculation of such *CAV*.

$$CAV_{j, current, t} = CAV_{j, current, t-1}[1 - \frac{(t - T_{start_j})}{(T_{end_j} - T_{start_j})}], \quad (6)$$

where $CAV_{j, current, t}$ and $CAV_{j, current, t-1}$ respectively represent the *CAV* of activity $j$ for participants involved at time $t$ and $t$-1, and $(t - T_{start_j})/(T_{end_j} - T_{start_j})$ is the progress of activity $j$. The factor of *ADR* is disregarded in this part to avoid a double impact on the calculation of the interest matching value.

The two aforementioned calculation methods require special settings for the conference personnel who need to participate in some activities passively, such as seminar chairs and keynote speakers. Their *CIV* and *CAV* values to corresponding activities are set at the maximum, and the factors affecting the changes in *CIV* and *CAV* values, such as *IDR*, *IRR*, *ADR*, and *ADT*, are set at 0. These settings ensure the participation of relevant participants in specific conference activities during the data generation process.

The subjective factors, which motivate each participant to decide the next movement (e.g., to stay at the current activity, to turn to the next activity, or to leave the venue), are simulated through the above calculations and settings. However, their behavioral decisions are also dictated by objective factors, for which the behavior-controlled model is designed.

*C. Behavior-controlled model*

The behaviors of people may change because of the external environment (e.g., opening of an activity) and internal interests (e.g., interest changes). A behavior-controlled model is proposed to simulate the conference schedule and control crowd movements as the schedule proceeds. This model comprises a four-part updating operation and six control strategies. The four-part updating operation aims to update the values of status attributes of conference personnel, conference activities, and indoor space units. The operation includes the following: (1) updating the status attributes and the energy value of each individual; (2) updating the status attributes of each activity; (3) updating the status attributes, saturation, and the current maximum moving speed of each space unit; (4) updating the interest and attracting values of each activity to each participant. The six control strategies ensure the reasonability of every moving decision-making as follows.

**Strategy 1**: Decision on admission time. This strategy determines the time of each day when the participants first enter the venue. The *CIV*s and *CAV*s of the crowd and all venue activities are initialized when the opening of the venue, the activity of the highest matching interest value is selected for each individual, and a slightly earlier or later time than the starting time of the activity is set as their entry time.

**Strategy 2**: Judgment of loss of interest. This strategy is employed to control the person who is present at an activity to decide to leave as their interest fades away. A global control constant, namely *LIV* (interest value for losing), is introduced. If the current interest matching value of an individual to their ongoing activity is lower than the *LIV*, then these individuals will tend to leave the activity and choose to turn to another available activity based on strategy 3.

**Strategy 3**: Decision of activity switching. This strategy traverses the interest matching values of the conference personnel who need to switch conference activities at the moment to all conference activities to determine their next moving destination. A global control constant, namely *SIV* (interest value for switching), is introduced to determine if the current interest matching value of an individual is sufficiently large to prompt participation in an available activity as a switching target. If multiple optional activities reaching the *MIV* value simultaneously are available, then the individual will switch to the activity with the highest interest matching value.

**Strategy 4**: Preemption of high-priority activities. This strategy is leveraged to control the people who are currently participating in a certain activity to decide leaving due to the start of another activity with high priority. If the current interest matching value of an individual to an available activity reaches the *MIV* value, the individuals will immediately turn away from the current activity for the new target activity based on the priority attribute of conference activities. For example, when a keynote whose priority is far above the exhibition starts, a participant will immediately stop watching and turn to the main venue when the participant is watching the exhibition and is still highly interested in the exhibition because his/her current interest matching value for the keynote has reached the *MIV*.

**Strategy 5**: Decision of exiting time. This strategy determines the exiting time of participants and mainly includes three assessment criteria: (1) the current energy value of a participant is less than the minimum; (2) the interest matching values for all activities do not reach the *MIV* value; (3) the current time is close to the venue closing time.

**Strategy 6**: Overall control of movements. The movements of participants will be constrained by the saturation and the maximum speed of each space unit when they enter or leave the venue or switch activities.

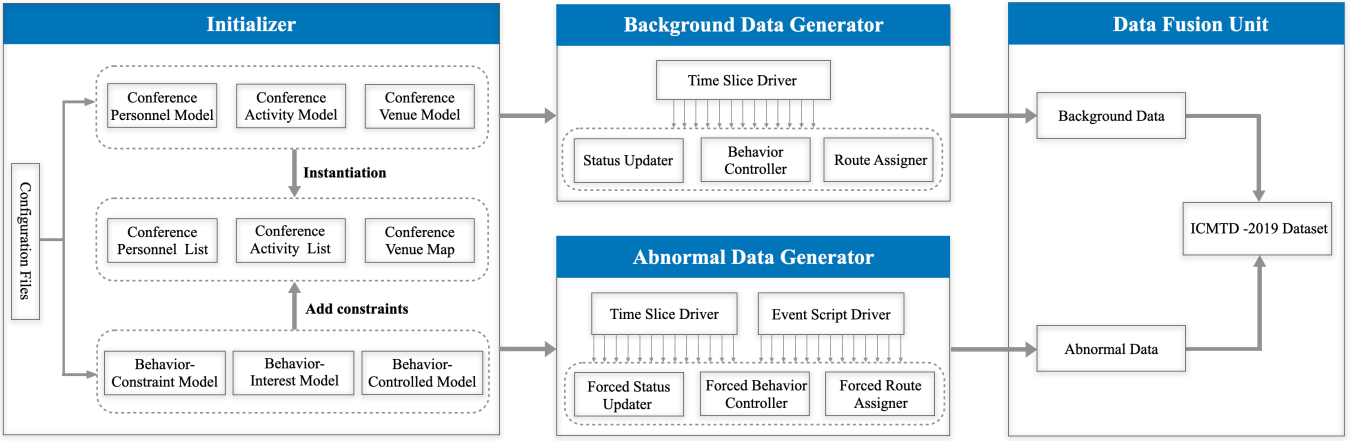

Fig. 2. Workflow of data generator

## VI. DATA GENERATION

A data generator of indoor crowd movement trajectories was designed in accordance with the results of entity and behavior modeling. The data generator was developed using the node.js development environment on a Linux computer system. Fig. 2 shows that the data generator comprises four modules, namely initializer, background data generator, abnormal data generator, and data fusion unit. The four modules are introduced as follows.

### A. Initialization

The initializer creates and initializes the three entity and three behavior models. Moreover, the initializer uses configuration files to set the attributes of the models. Its output is a profile file that includes a conference personnel list, a conference activity list, a conference venue map, and three initial behavioral models. The initialization process has the following main steps.

Step 1: Set data types and their data fields.

Step 2: Create and initialize conference personnel, conference activity, and conference venue models.

Step 3: Instantiate the conference personnel model, conference activity model, and the conference venue to generate a conference personnel list, activity list, and venue map, respectively.

Step 4: Create and initialize a behavior-constraint model. First, time and capacity constraints are added to the conference venue map. Then, an energy constraint based on the basic attributes of each individual in the conference personnel list is added. Last, a permission constraint is added to each activity in the conference activity list.

Step 5: Create a behavior interest model and initialize the interest of each individual in each activity and the attractiveness of each activity to each individual.

Step 6: Create and initialize a behavior-controlled model and set the three global control constants, namely LIV, SIV, and MIV.

Step 7: Integrate all the initialization results to generate a profile file, which can be further tuned manually. For example, we adjust the value of the attractiveness parameter associated with an abnormal activity to ensure it perform abnormally.

In the initialization, three groups of parameters need to be set. The first group of parameters (e.g., the types of people, the time and location of each conference activity, and the capacity of each grid), as listed in tables I, II, and III, is related to the three entity models. The second group is the quantity parameters (e.g., the number of persons to be generated) that are set during the entity instantiation. The third group is the behavior control parameters, such as LIV and SIV, set in the behavior models.

### B. Background data generation

The background data generator simulates the normal behaviors of all conference personnel during the conference to generate the background data. The output of this generator is a CSV file, which records the movements of normal behaviors of all conference personnel. The background data generator has four sub-modules, namely time slice driver, status updater, behavior controller, and route assigner, as shown in Fig. 2. The background data generation process has the following main steps.

Step 1: The time slice driver bins the entire opening time of the conference venue by seconds and simulates the time to go forward.

Step 2: The status updater for a certain time bin updates the various statuses and attribute values of conference personnel, activities, and space units.

Step 3: The behavior controller determines whether each of the conference personnel needs to switch conference activities in the time bin. The behavior controller creates the next target activity for each one who needs to switch based on the six moving decision-making strategies. We regard the venue map with grids as a graph in which a node represents a grid and an edge represents the relation between two adjacent grids. The route assigner adopts the shortest breadth-first path-finding algorithm [69] to allocate a path from the current activity to the target activity after considering the real-time saturation and speed constraints of each passing grid. Also, some random perturbations are added to the path to make the actual movements realistic.

Step 4: The time slice driver simulates the movements of conference personnel in the time bin. Then, the time slice driver starts a new time bin and returns to step 2.

TABLE VI
MOVEMENT PATTERNS OF SEVEN TYPES OF CONFERENCE PERSONNEL

| Type | Basic Movement Patterns |
|---|---|
| VIP guest | Takes the VIP channel for entry; no need to sign in; usually rests in the VIP lounge; mainly focuses on the activities in the main venue and sub-venues; often sits in the front row of the venue; |
| Ordinary guest | Sign-in required for entry; mainly moves in the main venue, sub-venues, exhibition halls, and poster areas; independently participates in conference activities based on personal interests; |
| Visitor | Similar to ordinary guests but not authorized to enter the main venue and unavailable to lunches and dinners provided by the hosts; |
| Media reporter | Sign-in for entry; rests and works in the media room; often goes to other areas to conduct live broadcasts and interviews; |
| Hacking contestant | Sign-in for entry; mainly moves in the hacking contest area; |
| Staff | Enters the venue in advance to be ready for work; distributed throughout the venue, with own fixed working places and scopes of responsibility; often enters and exits the work room and haves lunch in that area; |
| Exhibitor | Sign-in for entry; mainly moves in the exhibition area. |

## C. Abnormal data generation

Twelve abnormal events are defined in the scene (Section VII). An event script is used to describe the event storylines and record the involved persons, conference activities, and moving routes. The output of the generator is a CSV file that records the movements of these abnormal behaviors. The abnormal data generator has five sub-modules, namely time slice driver, event script driver, forced status updater, forced behavior controller, and forced route assigner, as shown in Fig. 2. These modules work together to generate data case by case for each abnormal event. The abnormal data generation process comprises the following main steps.

Step 1: The time slice and event script driver simulate the timeline of each abnormal event by time bin.

Step 2: The forced status updater for a certain event forcibly changes the status and attribute values of relevant persons and activities when the timeline of the event begins.

Step 3: The forced behavior controller for the events ensures that the relevant persons appear at the preset activity and time in the script. The forced route generator ensures that the relevant persons walk along with the predefined route when moving to the target activity.

## D. Data fusion

The data fusion unit integrates the background and abnormal data. Possible conflicts in the two types of data are observed in the data fusion. For example, a person related to an abnormal event may have his trajectories in the background and abnormal data at a certain time. The principle of prioritizing the retention of trajectories in abnormal events is adopted to eliminate such conflicts and ensure the smooth connections of the two types of data. The fused data are sorted in chronological order after the conflicts are eliminated. Moreover, when an individual focuses on a conference activity without movements, they will generate a number of trajectory points at the same position within a continuous period of time. Therefore, we remove such duplicate trajectory points and save the trajectories of individuals entering a position to reduce the size of the dataset and emphasize the movement behaviors of conference personnel in the venue, ultimately to obtain the final ICMTD-2019 dataset.

## VII. GROUND TRUTH

### A. Background

China Intelligence Cyber Security Conference is a fictitious academic conference in the field of intelligent cyber security. The conference, which lasted for three days, invited numerous senior experts to give keynotes and set-up a series of academic activities related to six popular topics, including data security, IoT security, mobile security, privacy protection, smart venue, and smart security technology innovation. The conference also hosted a large business exhibition and a hacking contest, combined with a variety of social activities, such as tea breaks, interviews, and a banquet. The detailed schedule of the conference is provided in the supplementary materials of this paper. Every participant and organizer wore a smart badge during the conference to capture their movements in the public areas of the venue.

### B. Conference personnel

A total of 5256 people, which can be classified into seven types of personnel with different venue permissions and movement patterns, attended the conference as shown in Table VI.

### C. Abnormal events

Twelve notable abnormal events were found during the conference. These events and involved figures are introduced as follows.

E1: Copy of name badge. Personnel A used a copy of the name badge of VIP guest B (PID-16632) and wore the copied badge to enter the VIP lounge, staying in the lounge for approximately one hour. Personnel A was suspected of theft.

E2: Item missing. VIP guest C (PID-11260) rested in the VIP lounge after lunch and frequently went to the service desk for inquiry of a possibly missed item. Personnel A was suspected to had stolen the item of VIP C because personnel A stayed in the VIP lounge for a long time and hurriedly left the venue after VIP C entered.

E3: Equipment failure. On the second day of the conference, sensors (SID-10715, 10716, 10717, 10718, 10815, 10816, 10817, and 10818) failed from 13:33:21 to 13:56:15, resulting in missing sensing data.

E4: Personnel ultra vires. The VIP lounge and the media room were dedicated to VIP guests and media reporters, respectively. Other types of personnel were not authorized to enter these areas. However, two media reporters (PID-11201 and 16473) acted beyond their authority to enter the VIP lounge at 9:00–9:20 and 10:30–10:50, respectively, on the first day of the conference; a VIP guest (PID-13344) entered the media room at 12:29:04–12:39:50 on the second day.

TABLE VII
IMPORTANCE, DISCOVERY RATE, ACCURACY AND DIFFICULT LEVEL OF ABNORMAL EVENTS

| Abnormal Event | Illustration | Importance | Discovery Rate | Accuracy | Difficulty |
|---|---|---|---|---|---|
| E1 | Copy of name badge | Important | 32% | 21.9% | Simple |
| E2 | Item missing | | 1.3% | 1.0% | Difficult |
| E3 | Equipment failure | | 12.0% | 8.7% | Medium |
| E4 | Personnel ultra vires | | 2.7% | 1.8% | Difficult |
| E5 | Packed sub-venues | | 6.7% | 6.2% | Medium |
| E6 | Venue congestion | Moderate | 44.0% | 21.0% | Simple |
| E7 | Book signing | | 24.0% | 19.2% | Medium |
| E8 | Group visit | | 8.0% | 3.3% | Difficult |
| E9 | Forgotten badge | | 2.7% | 1.7% | Difficult |
| E10 | Early exit of hacking contest | General | 5.3% | 4.7% | Difficult |
| E11 | Staff lateness | | 20.0% | 15.6% | Medium |
| E12 | Staff lunch turns | | 5.3% | 2.7% | Difficult |

E5: Packed sub-venues. Three sub-venue activities drawn the attention of unexpectedly numerous participants. These activities were the Internet of Things Security Forum held at Sub-venue B at 14:00–16:15 on the second day, the Mobile Security Forum held at Sub-venue B from 9:30–11:30 on the third day, and the Intelligent Security Technology Innovation Forum held in Sub-venue A at 10:30–11:30 on the third day.

E6: Venue congestion. The spaces of tea break areas, restrooms, and corridors were limited, and the staff failed to guide crowd movements effectively, resulting in congestions during tea breaks.

E7: Book signing. A book signing was held in the exhibition hall at 13:00–14:30 the second day, with a large number of participants gathering in the hall.

E8: Group visit. During the conference, four visiting groups toured the exhibition hall and poster area of the venue at 10:00–11:00, 15:00–16:00 on the first day and 10:00–11:00, 15:00–16:00 on the second day, respectively. Visiting groups generally acted collectively, with a size of approximately 100 people.

E9: Forgotten badge. VIP guest G (PID-19929) was the chair of the hacking contest. At 9:00 on the second day, VIP guest G came to the hacking contest venue and forgot the badge on the podium until the end of the contest around 17:30 when VIP guest G returned to the podium to get the badge back and then left the venue.

E10: Early exit of hacking contestants. The hacking contest was divided into two parts: basic and additional assessments. The basic assessment was conducted in the morning of the first and second days of the conference, whereas the additional assessments were held in the afternoon of the first and second days and the morning of the third day, adopting a knockout system. Thus, some contestants left the venue soon after being eliminated.

E11: Staff lateness. Staff should enter the venue in advance. However, some staff, including: PID-18347, 10345, 14859, 18059, 12856, 11396, 14678, 10762, and 17576, were late.

E12: Staff lunch turns. The staff in the venue were divided into two groups to have lunch alternately at the work room, and their dining times were 11:40–12:10 and 12:10–12:40.

## VIII. EVALUATION

A two-phase evaluation was conducted to verify the completeness, usability, and validity of the ICMTD-2019 dataset. The first phase was an internal test, in which an independent test team was invited to test the dataset completeness and analyze the movement patterns of various groups of people and abnormal events without knowing the ground truth. The test results showed that the ICMTD-2019 dataset has good completeness and usability and it behaves consistently with the ground truth. The second phase was an external test, in which the ICMTD-2019 dataset was used by ChinaVis Data Challenge 2019 [54]. The external test will be comprehensively discussed below.

### A. Evaluation process

ChinaVis Data Challenge 2019 invited researchers, developers, and amateurs who used their most effective methods, techniques, and tools to analyze the ICMTD-2019 dataset. Their data analytics tasks included the following: (1) inferring the schedule of the conference; (2) analyzing the types of conference personnel in the venue and summarizing the movement patterns of each type; (3) identifying at least five abnormal events; (4) summarizing the deficiencies in conference organization and management. Contestants were required to submit analysis results and demonstration videos.
The organizing committee of the data challenge invited experts of public security and visual analytics to review the entries jointly. Each entry was randomly assigned to four to six experts who gave scores and comments from the five aspects of analysis quality, visual design, interaction design, originality, and scalability according to the ground truth and personal experience. The entry score was based on a five-point system, with five being the best and one the worst. The supplementary materials of this paper provide the introduction of some entries. Three types of feedback in the course of the data challenge were collected to evaluate the effectiveness of the ICMTD-2019 dataset: (1) expert review comments and scores; (2) discovery rate and accuracy of the abnormal event analysis of the entries;

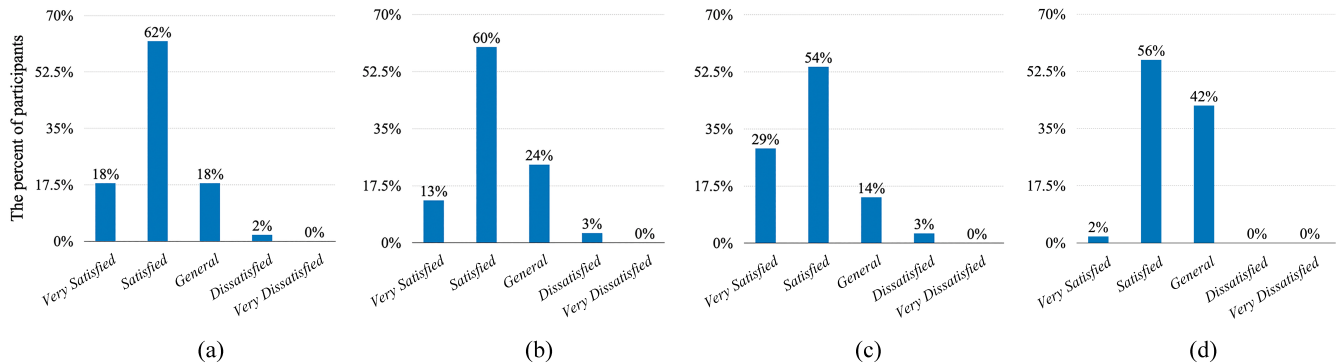

Fig. 3. Questionnaire results from the contestants of China Vis Data Challenge 2019. (a) Scene design satisfaction; (b) Dataset overall quality satisfaction; (c) Review results satisfaction; (d) Task setting difficulty satisfaction.

(3) questionnaire of the contestants. The evaluation results will be presented below.

### B. Evaluation results

#### 1) Analysis quality evaluation

A total of 75 entries from 359 contestants were received. The expert review results indicate that the average score of the entries was 2.7 points, and the average variance was 0.5 points. Among these results, 32 entries were scored above 2.7 points, mainly concentrating between 3.0–3.5 points. The average scores of the five aspects (i.e., analysis quality, visual design, interaction design, originality, and scalability) were 2.7, 2.8, 2.7, 2.6, and 2.6, respectively. The results reveal that a majority of contestants completed most of the data analytics tasks. The overall difficulty of the ICMTD-2019 dataset was moderate, and data users can effectively analyze most of the preset movement patterns and abnormal events in the data.

In addition, we invited domain experts to assess the importance of each abnormal event according to the severity of threat it posed to venue management. The 12 abnormal events were divided into 4 important events, 4 moderate events and 4 general events. Important events (E1–E4) usually brought severe threats to venue security or personal safety. Moderate events (E5–E8) usually led to crowding in the venue and therefore may put the safety of persons at risk. General events (E9–E12) did not directly affect persons and activities in the venue.

#### 2) Event difficulty evaluation

The discovery rate and accuracy of the 12 abnormal events were calculated for each entry. The discovery rate was the ratio of the number of entries from which an abnormal event was found to the total number of entries. The accuracy was the average completeness of the information concerning abnormal event identification (i.e., time, locations, persons, and event contents) of all entries. The results indicate that the difficulty of correct analysis of each event was depicted on the basis of the accuracy indicator. The events with an accuracy larger than 20%, between 5.0% and 20%, and less than 5.0% were simple, medium-difficult, and difficult events, respectively. Table VII shows that the 12 abnormal events were classified into two simple, four medium-difficult, and six difficult events. Consequently, the difficulty of the abnormal events embedded in the ICMTD-2019 dataset was varied and reasonable. Simple and medium-difficult events can be found in many entries. Several high-difficulty events were also found to challenge the contestants.

Simple events obtained relatively high discovery rates and accuracies, with distinctive crowd movement patterns. The contestants can identify abnormal events by observing the overall distributions of trajectories in the venue, such as discontinuous or repeated trajectories (E1) and large-scale crowd gathering (E6). For moderately difficult events, the contestants must analyze crowd movements and their distributions from a fine-gained time and space granularity, such as missing data (E3) and gatherings of small groups of people (E5 and E7). Moreover, the contestants must to correctly identify personnel types with unique movement trajectories, such as the lateness of some staff (E11). The analysis challenges of difficult events originate from two aspects. (1) The contestants must correctly and completely identify the permissions and movement patterns of all types of conference personnel and then recognize the VIP lounge, media room, hacking contest area, tea break area 1, tea break area 2, and work room that were not disclosed in the map provided in the data challenge. For example, E4 required contestants to identify the media room and VIP lounge, as well as the corresponding media reporters and VIP guests who were allowed to enter. (2) The contestants were expected to determine sudden changes of movement patterns and combine them with other abnormal events, including E2, E9, and E10, to explore the hidden truth jointly. For instance, E2 required recognition of the frequent movements of an individual between the VIP lounge and the service during a short time and combination of the clue of E1.

#### 3) Subjective evaluation

The subjective feedback of the contestants on the scene design of the dataset, the overall quality of the dataset, the results of the entry review, and the difficulty of analysis tasks were collected in the form of questionnaires. A total of 55 valid answers were received, and the results are shown in Fig. 3. Approximately 80% of the contestants were satisfied with the scene design of the dataset (Fig. 3(a)). These contestants believed that the scene was realistic, the background was completed, and the design of characters, events, and spatial structure was reasonable. Approximately 73% of these contestants commented that the overall quality of the ICMTD-2019 dataset was good (Fig. 3 (b)). Moreover, the dataset size and time span were moderate, containing many interesting clues and rich temporal and spatial patterns of crowd movements, which can elicit the enthusiasm of people for analysis. Approximately 83% expressed satisfaction with review scores and comments (Fig. 3(c)), which showed the fairness and rationality of expert views of the entries. Approximately 58% believed that the analytics tasks were slightly challenging (Fig. 3(d)), which can identify the effectiveness of the methods, technologies, and systems.

## IX. Conclusion

This paper introduces a benchmark dataset of crowd movement trajectories in smart venues. The scene design, modeling process, and generation method of the dataset and its ground truth are elaborated. In addition, the application of the ICMTD-2019 dataset in ChinaVis Data Challenge 2019 is comprehensively presented to verify the completeness, usability, and validity of the dataset. The results show that the ICMTD-2019 dataset presents good completeness and usability, and can effectively identify the performance of methods, technologies, and systems for indoor trajectory analysis.

This work still has some limitations. First, in terms of the scalability of the data generator, on the one hand, all parameters of the proposed data generator can be adjusted, making it convenient to generate new data. Specially, an arbitrary number of participants can be set to generate new data of different sizes. On the other hand, the scene and the map of our data generator cannot support major changes that may result in the redesign of the entity and behavior models, which is the main limitation of this work.

Second, the three behavior models are relatively simple, leading to difficulties in achieving a complete representation of the complexity and diversity of human behaviors in the real world. A lot are worth doing to improve the behavior modeling. For example, the energy of a person in a venue can be recharged briefly by taking a short break.

Finally, a square area of $8 \times 8 \text{ m}^2$ is considered to be a space unit when modeling the venue space, but such division is still of relatively large granularity, making it hard to reflect high-precision movement patterns.

In future work, we will be devoted to designing data generation methods with less constraints by the scenario. We will take into account more human factors in order to simulate realistic and diverse human behaviors. Furthermore, high-precision indoor spaces will be considered to produce subtle movement patterns and characteristics. At last, the combination of data- and program-driven methods should be a good direction for benchmark data simulation.

## Acknowledgment

Thanks to all the organizers, reviewers, and participants of ChinaVis Data Challenge 2019. Thanks to Qi Ma, Xueshi Wei and Chenhua Xie from Qi An Xin Technology Group Inc. for their fruitful discussions. The work is supported in part by the National Natural Science Foundation of China (No.61872388 and 62072470). The ICMTD-2019 dataset is available at https://github.com/csuvis/IndoorTrajectoryData/ and http://www.chinavis.org/2019/english/challenge_en.html.